\documentclass[twocolumn,longbib]{aastex7}

\usepackage{CJK}
\usepackage{fontawesome}

\hypersetup{pdfauthor={Name}}

\newcommand{\githubicon}{{\color{black}\faGithub}}

\begin{document}
\begin{CJK*}{UTF8}{gbsn}

\title{Discovery and Preliminary Characterization of a Third Interstellar Object: 3I/ATLAS}

\author[orcid=0000-0002-0726-6480]{Darryl Z. Seligman}
\altaffiliation{NSF Astronomy and Astrophysics Postdoctoral Fellow}
\affiliation{Department of Physics and Astronomy, Michigan State University, East Lansing, MI 48824, USA}
\email[show]{dzs@msu.edu}  

\author[0000-0001-7895-8209]{Marco Micheli}
\affiliation{ESA NEO Coordination Centre, Largo Galileo Galilei 1,
 I-00044 Frascati (RM), Italy}
\email{marco.micheli@ext.esa.int}  

\author[0000-0003-0774-884X]{Davide Farnocchia}
\affiliation{Jet Propulsion Laboratory, California Institute of Technology,
4800 Oak Grove Dr., Pasadena, CA 91109, USA}
\email{davide.farnocchia@jpl.nasa.gov}  

\author[0000-0002-7034-148X]{Larry Denneau}
\affiliation{Institute for Astronomy, University of Hawaii, 2680 Woodlawn Drive, Honolulu, HI 96822, USA}
\email{denneau@hawaii.edu}

\author[0000-0003-2152-6987]{John W.\ Noonan}
\affiliation{Department of Physics, Auburn University, Edmund C.\ Leach Science Center, Auburn, 36849, AL, USA}
\email{noonan@auburn.edu}

\author[0000-0001-7225-9271]{Henry H. Hsieh}
\affiliation{Planetary Science Institute, 1700 East Fort Lowell Rd., Suite 106, Tucson, AZ 85719, USA}
 \email{hhsieh@psi.edu}

\author[0000-0002-0143-9440]{Toni Santana-Ros}
\affiliation{Departamento de F\'{\i}sica, Ingenier\'{\i}a de Sistemas y Teor\'{\i}a de la Se\~{n}al, Universidad de Alicante, Carr. San Vicente del Raspeig, s/n, 03690 San Vicente del Raspeig, Alicante, Spain}
\affiliation{Institut de Ci\`encies del Cosmos (ICCUB), Universitat de Barcelona (UB), c. Mart\'i Franqu\`es, 1, 08028 Barcelona, Catalonia, Spain}
\email{tsantanaros@icc.ub.edu}  

 \author[0000-0003-2858-9657]{John Tonry}
\affiliation{Institute for Astronomy, University of Hawaii, 2680 Woodlawn Drive, Honolulu, HI 96822, USA}
\email{tonry@hawaii.edu}

%\author{K.~Auchettl$^{6,3}$\orcidlink{0000-0002-4449-9152}}
\author[0000-0002-4449-9152]{Katie Auchettl}
\affiliation{School of Physics, University of Melbourne, Parkville, VIC 3010, Australia}
 \affiliation{Department of Astronomy and Astrophysics, University of California, Santa Cruz, CA 93105, USA}
 \email{katie.auchettl@unimelb.edu.au} 

 \author[0000-0002-6710-8476]{Luca Conversi}
\affiliation{ESA NEO Coordination Centre, Largo Galileo Galilei 1,
 I-00044 Frascati (RM), Italy}
 \email{luca.conversi@esa.int}

\author[0000-0002-6509-6360]{Maxime Devog\`ele}
\affiliation{ESA NEO Coordination Centre, Largo Galileo Galilei 1,
 I-00044 Frascati (RM), Italy}
 \email{maxime.devogele@ext.esa.int}  

\author[0000-0002-5447-432X]{Laura Faggioli}
\affiliation{ESA NEO Coordination Centre, Largo Galileo Galilei 1,
 I-00044 Frascati (RM), Italy}
 \email{laura.faggioli@ext.esa.int} 

\author[0000-0002-9464-8101]{Adina~D.~Feinstein}
\altaffiliation{NHFP Sagan Fellow}
\affiliation{Department of Physics and Astronomy, Michigan State University, East Lansing, MI 48824, USA}
\email{adina@msu.edu}

\author[0000-0002-7058-0413]{Marco Fenucci}
\affiliation{ESA NEO Coordination Centre, Largo Galileo Galilei 1,
 I-00044 Frascati (RM), Italy}
 \email{marco.fenucci@ext.esa.int} 

 \author[0000-0002-0535-652X]{Marin Ferrais}
\affiliation{Florida Space Institute, University of Central Florida, 12354 Research Parkway, Orlando, FL 32828, USA}
 \email{marin.ferrais@ucf.edu}  

\author[0009-0000-4697-5450]{Tessa Frincke}
\affiliation{Department of Physics and Astronomy, Michigan State University, East Lansing, MI 48824, USA}
\email{frincket@msu.edu}

\author[0000-0003-1462-7739]{Michael Gillon} 
\affiliation{Astrobiology Research Unit, Universit\'e de Li\`ege, 4000 Li\`ege, Belgium}
 \email{Michael.Gillon@uliege.be}

\author[0000-0001-6952-9349]{Olivier R. Hainaut}
\affiliation{European Southern Observatory, Karl-Schwarzschild-St. 2, 85748 Garching-bei-M\"unchen, Germany}
\email{ohainaut@eso.org}

\author{Kyle Hart}
\affiliation{Institute for Astronomy, University of Hawaii, 
2680 Woodlawn Drive, Honolulu, HI 96822, USA}
\email{kylehart@hawaii.edu} 

\author[0000-0002-8732-6980]{Andrew Hoffman}
\affiliation{Institute for Astronomy, University of Hawaii, 
2680 Woodlawn Drive, Honolulu, HI 96822, USA}
\email{amhoh@hawaii.edu} 

\author[0000-0002-4043-6445]{Carrie E.\ Holt}
\affiliation{Las Cumbres Observatory, 6740 Cortona Drive, Suite 102, Goleta, CA 93117, USA}
\email{cholt@lco.global}

\author[0000-0003-3953-9532]{Willem~B.~Hoogendam}
\altaffiliation{NSF Graduate Research Fellow}
\affiliation{Institute for Astronomy, University of Hawaii, 
2680 Woodlawn Drive, Honolulu, HI 96822, USA}
\email{willemh@hawaii.edu}  

\author[0000-0003-1059-9603]{Mark~E.~Huber}
\affiliation{Institute for Astronomy, University of Hawaii, 
2680 Woodlawn Drive, Honolulu, HI 96822, USA}
\email{mehuber7@hawaii.edu} 

\author[0000-0001-8923-488X]{Emmanuel Jehin}
\affiliation{Space sciences, Technologies \& Astrophysics Research (STAR) Institute Universit\'e de Li\'ege 4000 Liege, Belgium}
 \email{ejehin@uliege.be}

\author[0000-0003-1008-7499]{Theodore Kareta}
\affiliation{Dept. of Astrophysics and Planetary Science, Villanova University, Villanova, PA, USA}
\affiliation{Lowell Observatory, 1400 W Mars Hill Rd, Flagstaff, AZ 86001, USA}
\email{theodore.kareta@villanova.edu}

\author[0000-0002-2021-1863]{Jacqueline V. Keane}
\affiliation{U.S. National Science Foundation, 2415 Eisenhower Avenue, Alexandria, VA 22314, USA}
\email{jvkeane@hawaii.edu}

\author[0000-0002-6702-7676]{Michael S.\ P.\ Kelley}
\affiliation{Department of Astronomy, University of Maryland, College Park, MD 20742-0001, USA}
\email{msk@astro.umd.edu}

\author[0000-0002-3818-7769]{Tim Lister}
\affiliation{Las Cumbres Observatory, 6740 Cortona Drive, Suite 102, Goleta, CA 93117, USA}
\email{tlister@lco.global}

\author[0000-0001-8397-3315]{Kathleen Mandt}
\affiliation{NASA Goddard Space Flight Center, Greenbelt, MD, 20771, USA}
\email{kathleen.mandt@nasa.gov}

\author[0000-0002-6930-2205]{Jean Manfroid}
\affiliation{Space sciences, Technologies \& Astrophysics Research (STAR) Institute Universit\'e de Li\'ege 4000 Liege, Belgium}
 \email{jmanfroid@gmail.com}

\author[0000-0003-4706-4602]{Du{\v s}an Mar{\v c}eta}
\affiliation{Department of Astronomy, Faculty of Mathematics, University of Belgrade,  Serbia}
\email{dusan.marceta@matf.bg.ac.rs}

\author[0000-0002-2058-5670]{Karen J. Meech}
\affiliation{Institute for Astronomy, University of Hawaii, 
2680 Woodlawn Drive, Honolulu, HI 96822, USA}
\email{meech@hawaii.edu}  

\author{Mohamed Amine Miftah}
\affiliation{Space sciences, Technologies \& Astrophysics Research (STAR) Institute Universit\'e de Li\'ege 4000 Liege, Belgium}
\affiliation{Cadi Ayyad University (UCA), Oukaimeden Observatory (OUCA), Faculté des Sciences Semlalia (FSSM), High Energy Physics, Astrophysics and Geoscience Laboratory (LPHEAG), Marrakech, Morocco}
\email{mohamedamine.miftah@uliege.be}  

\author[0000-0003-4022-6234]{Marvin Morgan}
\affiliation{Department of Physics, University of California, Santa Barbara, Santa Barbara, CA 93106, USA}
\email{marvinmorgan@ucsb.edu}  

\author[0000-0002-9836-3285]{Francisco Oca\~na}
\affiliation{ESA NEO Coordination Centre, Largo Galileo Galilei 1,
 I-00044 Frascati (RM), Italy}
\email{francisco.ocana@ext.esa.int} 

\author[0000-0002-7257-2150]{Eloy Pe\~na-Asensio}
\affiliation{Department of Aerospace Science and Technology, Politecnico di Milano, Via La Masa 34, 20156 Milano, Italy}
\email{eloy.pena@polimi.it}  

\author[0000-0003-4631-1149]{Benjamin~J.~Shappee}
\affiliation{Institute for Astronomy, University of Hawaii, 
2680 Woodlawn Drive, Honolulu, HI 96822, USA}
\email{shappee@hawaii.edu}

\author[0000-0001-5016-3359]{Robert J. Siverd}
\affiliation{Institute for Astronomy, University of Hawaii, 
2680 Woodlawn Drive, Honolulu, HI 96822, USA}
\email{rsiverd@hawaii.edu}

\author[0000-0002-0140-4475]{Aster G. Taylor}
\altaffiliation{Fannie and John Hertz Foundation Fellow}
\affiliation{Dept. of Astronomy, University of Michigan, Ann Arbor, MI 48109, USA}
\email{agtaylor@umich.edu}

\author[0000-0002-2471-8442]{Michael~A.~Tucker}
\affiliation{Center for Cosmology \& Astroparticle Physics, The Ohio State University, Columbus, OH, USA}
\affiliation{Department of Astronomy, The Ohio State University, Columbus, OH, USA}
\email{tucker.957@osu.edu}

\author[0000-0002-1341-0952]{Richard Wainscoat}
\affiliation{Institute for Astronomy, University of Hawaii, 2680 Woodlawn Drive, Honolulu, HI 96822, USA}
\email{rjw@hawaii.edu}

\author[0000-0002-0439-9341]{Robert Weryk}
\affiliation{Department of Physics and Astronomy, The University of Western Ontario, 1151 Richmond Street, London, ON N6A 3K7, Canada}
\email{rweryk@uwo.ca}

\author[0000-0001-5559-2179]{James J. Wray}
\affiliation{School of Earth and Atmospheric Sciences, Georgia Institute of Technology, 311 Ferst Drive, Atlanta, GA 30332, USA}
\email{jwray@gatech.edu}

\author[0009-0001-9538-1971]{Atsuhiro Yaginuma} 
\affiliation{Dept. of Physics and Astronomy, Michigan State University, East Lansing, MI 48824, USA}
\email{yaginuma@msu.edu}

\author[0000-0002-5033-9593]{Bin Yang}
\affiliation{Instituto de Estudios Astrof\'isicos, Facultad de Ingenier\'ia y Ciencias, Universidad Diego Portales, Santiago, Chile}
\email{bin.yang@mail.udp.cl} 

\author[0000-0002-4838-7676]{Quanzhi Ye (叶泉志)}
\affiliation{Department of Astronomy, University of Maryland, College Park, MD 20742-0001, USA}
\affiliation{Center for Space Physics, Boston University, 725 Commonwealth Ave, Boston, MA 02215, USA}
\email{qye@umd.edu}

\author[0000-0002-6702-191X]{Qicheng Zhang}
\affiliation{Lowell Observatory, 1400 W Mars Hill Rd, Flagstaff, AZ 86001, USA}
\email{qicheng@cometary.org}

\begin{abstract}
We report initial observations aimed at the characterization of a third interstellar object. This object, 3I/ATLAS or C/2025 N1 (ATLAS), was discovered on 2025 July 1 UT and has an orbital eccentricity of $e\sim6.1$, perihelion of $q\sim 1.36$ au, inclination of $\sim175^\circ$, and hyperbolic velocity of $V_\infty\sim 58$ km s$^{-1}$.  We report deep stacked images obtained using the Canada-France-Hawaii Telescope and the Very Large Telescope that resolve a compact coma. Using images obtained from several smaller ground-based telescopes, we find minimal light curve variation for the object over a $\sim4$ day time span. The visible/near-infrared spectral slope of the object is 17.1$\pm$0.2  \%/100 nm, comparable to other interstellar objects and primitive solar system small bodies (comets and D-type asteroids). 3I/ATLAS will be observable through early September 2025, then unobservable by Earth-based observatories near perihelion due to low solar elongation. It will be observable again from the ground in late November 2025. Although this limitation unfortunately prohibits detailed Earth-based observations at perihelion when the activity of 3I/ATLAS is likely to peak, spacecraft at Mars could be used to make valuable observations at this time. 
\end{abstract}

\keywords{\uat{Asteroids}{72} --- \uat{Comets}{280} --- \uat{Meteors}{1041} --- \uat{Interstellar Objects}{52}}

\section{Introduction}\label{sec:intro}

\setcounter{footnote}{0}

The first two interstellar objects identified traversing the inner Solar System, 1I/`Oumuamua \citep{Williams17} and 2I/Borisov \citep{borisov_2I_cbet}, were discovered in 2017 and 2019, respectively. It has been suggested that interstellar objects formed in protostellar disks \citep{Fitzsimmons2024} or the cores of giant molecular clouds \citep{Hsieh2021}. Although there is little hope of identifying the exact home system for a given interstellar object \citep{Hallatt2020}, they provide the best opportunity to directly measure the properties of small bodies that formed outside of our solar system.

These first interstellar objects displayed divergent properties. For one, 1I/`Oumuamua displayed no visible activity \citep{Meech2017,Ye2017,Jewitt2017,Trilling2018}, yet nongravitational acceleration was detected in its trajectory at the $\sim30\sigma$ level \citep{Micheli2018}. Meanwhile, 2I/Borisov displayed clear outgassing and dust ejection activity \citep{Jewitt2019b,Fitzsimmons:2019,Ye:2019,McKay2020,Guzik:2020,Hui2020,Kim2020,Cremonese2020,yang2021}. The excess velocity of 1I/`Oumuamua and 2I/Borisov also differed significantly. These objects had velocities of $V_\infty\sim26$ km s$^{-1}$ and $V_\infty\sim32$ km s$^{-1}$ respectively, which approximately correspond to ages of $\sim10^2$ and $\sim10^3$~Myr \citep{Mamajek2017,Gaidos2017a, Feng2018,Fernandes2018,Hallatt2020,Hsieh2021}. In addition, 1I/`Oumuamua displayed brightness variations of $\sim3.5$ magnitudes corresponding to an extreme oblate $6:6:1$ geometry \citep{Meech2017,Knight2017,Fraser2017,Belton2018,Mashchenko2019,Taylor2023} and had a moderately red reflectance spectrum \citep{Meech2017,Fitzsimmons2017,Ye2017}.

These divergent properties led to a variety of hypotheses regarding the provenance of the population. Although 2I/Borisov's coma was found to contain volatile species typically seen in comets \citep{Opitom:2019-borisov, Kareta:2019, lin2020,Bannister2020,Xing2020,Bagnulo2021,Aravind2021}, it had a high enrichment of CO relative to H$_2$O \citep{Bodewits2020, Cordiner2020}. These ratios differentiate its composition from most solar system comets, which are typically rich in H$_2$O and contain CO between 1--15\% relative to water \citep{2024come.book..459B}. 

1I/`Oumuamua's lack of visible activity despite its nongravitational acceleration has led to a variety of hypothesized origins. \citet{Micheli2018} noted that for radiation pressure to cause the nongravitational acceleration, the object must either have an exceptionally low density or an extreme geometry. Such a density is a possible byproduct of diffusion-limited aggregation formation processes in the outskirts of a protostellar disk \citep{MoroMartin2019}. Somewhat counterintuitively, it has also been demonstrated that such hypothetical structures bound together by weak van der Waals forces could survive tidal disruption from the Solar gravity \citep{Flekkoy19}. It was also hypothesized that such a fractal aggregate could form in the coma of an undiscovered parent interstellar comet \citep{Luu20}.  

Alternatively, 1I/`Oumuamua could have been outgassing volatiles with low levels of dust production, rendering it photometrically inactive in all extant observations \citep{Micheli2018,Sekanina2019,Seligman2020,Levine2021,Levine2021_h2,desch20211i,jackson20211i,Desch2022,Bergner2023}. This argument has been recently bolstered; since the discovery of 1I/`Oumuamua, a series of photometrically inactive near-Earth objects (NEOs) have been reported to have significant comet-like nongravitational accelerations \citep{Farnocchia2023,Seligman2023b,Seligman2024PNAS}. These objects imply that 1I/`Oumuamua-like nongravitational accelerations may be more common than previously thought \citep{Taylor2024}. Regardless, our understanding of interstellar objects is incomplete. See \citet{Fitzsimmons2024}, \citet{Seligman2023}, \citet{Jewitt2023ARAA}, and \citet{MoroMartin2022} for recent reviews on this topic.

In this paper, we report early observations of 3I/ATLAS \added{\citep{Denneau2025}}, the third interstellar object to be discovered after 1I/`Oumuamua and 2I/Borisov,  to help inform and coordinate follow-up observations. \added{Preliminary observations of 3I/ATLAS were reported contemporaneously with this paper \citep{Jewitt2025,Alarcon2025,Opitom2025,Belyakov2025, Kareta2025, Marcos2025,Yang2025}, along with theoretical calculations \citep{Taylor2025,loeb2025,Hopkins2025b,Hibberd2025,Yaginuma2025}. We provide Python scripts and the data behind several figures, which are denoted by the \githubicon\ icon in the figure captions.}

\section{Discovery and hyperbolic orbit characterization}\label{sec:orbit}

\subsection{Discovery}

\begin{figure}
    \centering
    \includegraphics[width=1.\linewidth]{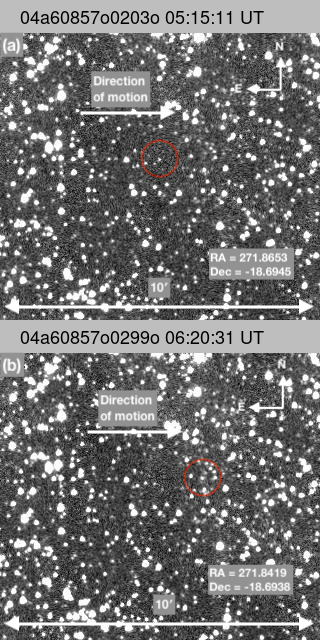}
    \caption{Cutout images from the first and fourth discovery observations of 3I/ATLAS from the ATLAS Chile, spanning approximately one hour. 3I/ATLAS is moving at 0.49 deg/day against the stellar background. The cardinal directions and direction of motion are indicated with arrows, and 3I/ATLAS is identified within the red circle. (a) Un-background subtracted image from 05:15:11 UT; (b) Un-background subtracted image from 06:20:31 UT. }
    \label{fig:discovery}
\end{figure}

3I/ATLAS was discovered through the robotic observing schedule from ATLAS Chile \citep{Tonry2018a} on 2025 July 1 and given the internal designation A11pl3Z\footnote{\url{https://www.minorplanetcenter.net/mpec/K25/K25N12.html}} (see Figure \ref{fig:discovery}). The discovery tracklet was immediately submitted to the Minor Planet Center (MPC). Follow-up observations were then conducted by ATLAS in Hawai`i, Sutherland, and the Canary Islands, along with dozens of other observatories worldwide. The discovery was made by ATLAS largely because the object was located in the Galactic plane --- a region typically avoided by more sensitive surveys, such as Pan-STARRS and the Catalina Sky Survey.

Six hours after the initial detection by ATLAS on 2025 July 1, pre-discovery detections from 2025 June 28--29 were identified in Zwicky Transient Facility \citep[ZTF;][]{bellm2019_ztf, 2019PASP..131g8001G} survey data, extending the orbital arc from 3.3 hours to 3 days. The new 3-day arc suggested a strongly hyperbolic orbit and prompted speculation on community mailing lists that the object might be interstellar. Later during the early hours of 2025 July 2 UT, additional pre-discovery pairs and triplets were identified in ATLAS data from five days earlier by S. Deen and later refined and resubmitted to the MPC by the ATLAS team. More pre-discovery detections from ZTF, dating back to 2025 June 14\footnote{Another batch of ZTF pre-discovery detections, dating back to 2025 May 22, were later submitted and published in MPEC 2025-N51 (\url{https://minorplanetcenter.net/mpec/K25/K25N51.html}).}, were also identified and submitted.

At the time of discovery in the ATLAS data, the object had a magnitude 17.7--17.9 in the $o$-band filter (see Figure \ref{fig:discovery}), and was at a heliocentric distance of $r=4.51$~au and a geocentric distance of $\Delta=3.50$~au.
\begin{figure}
    \centering
    \includegraphics[width=1.\linewidth, trim=3.5mm 3mm 4mm 3mm, clip]{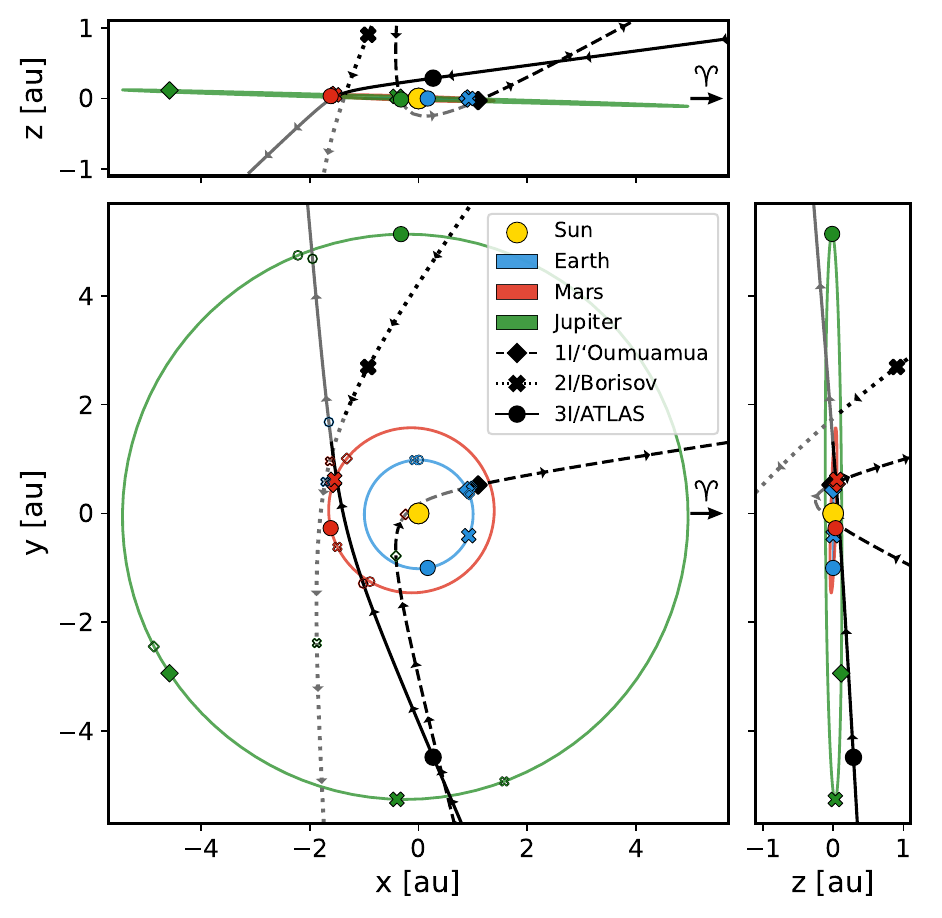}
    \caption{Heliocentric orbit (ECLIPJ2000) of 1I/`Oumuamua, 2I/Borisov, 3I/ATLAS, Earth, Mars, and Jupiter. Black lines represent the orbital path of each object; gray lines represent when the interstellar object is below-ecliptic. We highlight the location of the Sun (yellow) and the orbital paths of Earth (blue), Mars (red), and Jupiter (green). Large filled markers represent where the respective colored planet was when the interstellar object was discovered. Small unfilled markers represent where the planet will be at the interstellar object's closest approach. Arrows point along the motion direction. The vernal equinox is indicated to the right.}
    \label{fig:orbit}
\end{figure}

\subsection{Orbit Characterization}
As of the early hours of 2025 July 2, the initial discovery and pre-discovery arc, covering a total of about 18 days, was sufficient to confirm the large eccentricity and, consequently, hyperbolic nature of the object's orbit. As soon as the unusual nature of the object became evident, a large number of follow-up observations were obtained on 2025 July 2 by various observatories, leading to additional orbit refinement.

We first acquired 31 unfiltered 163~s exposures starting at 06:34 UT on 2025 July 2 spanning a total of 93 minutes using the European Space Agency's (ESA) 0.56-meter Test Bed Telescope (TBT; MPC code W57) at La Silla Observatory in Chile, which has a field of view of 2.5$^{\circ}$ $\times$ 2.5$^{\circ}$. The telescope is fully devoted to NEO survey and follow-up observations, and can be interrupted at any time for high-profile targets.

We also obtained detections with one of the Las Cumbres Observatory (LCO) 0.35-meter telescopes on Haleakal${\rm\bar{a}}$ (MPC code T03) and with the 2.0-meter Faulkes Telescope North (FTN; MPC code F65) and South (FTS, MPC code E10) between 2025 July 2 and July 4, which provided astrometric measurements with an astrometric accuracy better than $\pm0\farcs25$. 

A successful DDT request to the European Southern Observatory's (ESO) 8.2-meter Very Large Telescope (VLT), submitted on the same day and quickly approved, also allowed our team to obtain images of the object during good seeing on 2025 July 4. High-precision astrometry (better than $\pm0\farcs1$) was extracted from this dataset, using an extrapolation to zero aperture to ensure correction for possible asymmetries in the inner coma of the object. \added{\textit{Gaia} DR3 \citep{gaiacollaboration2023_gaiadr3} was used as the reference catalog for the astrometric solutions.}

Our best estimate of the heliocentric orbital elements of the object, at the time of submission of this paper, is listed in Table \ref{table:orbit}. In Figure \ref{fig:orbit}, we show the orbit of 3I/ATLAS in comparison to previously discovered interstellar objects.

\begin{table}
\centering
\caption{Initial orbit of 3I/ATLAS, computed using astrometry from 2025 May 22 to 2025 July 6. Heliocentric orbital elements at the epoch of 2025 July 3.}
\begin{tabular}{lc}
\hline
Orbital element & Value $\pm$ 1$\sigma$ \\
\hline
Perihelion distance $q$ [au]                    &   1.3558541 $\pm$ 0.0019855 \\
Eccentricity $e$                                &   6.1329072 $\pm$ 0.0162018 \\
Inclination $i$ [$^\circ$]                      & 175.11239   $\pm$ 0.00089  \\
Longitude of ascending node $\Omega$ [$^\circ$] & 322.14485   $\pm$ 0.01806  \\
Argument of perihelion $\omega$ [$^\circ$]      & 128.01038   $\pm$ 0.02340  \\
Time of perihelion $T_P$ [MJD, TDB]             & 60977.51085 $\pm$ 0.04078  \\
\hline
\end{tabular}
\label{table:orbit}
\end{table}

From the current heliocentric orbit, it is possible to infer the incoming trajectory of the object before it entered our Solar System. The eccentricity of the object's orbit with respect to the Solar System barycenter, computed before interacting with our planetary system, can be extrapolated as $e_b = 6.144 \pm 0.016$. This value, together with an incoming pericenter distance of $q_b = (1.3611 \pm 0.0020)$ au, results in an incoming velocity $v_{\infty} = (57.942 \pm 0.049)$ km s$^{-1}$, from an asymptote directed towards a Right Ascension of $\sim 295^\circ$ and a Declination of $\sim -19^\circ$, in the constellation of Sagittarius and not far from the Galactic Center.

\section{Observational Analysis}\label{sec:observations}

\subsection{Activity}\label{sec:activity}

\begin{figure}
    \includegraphics[width=1.\linewidth]{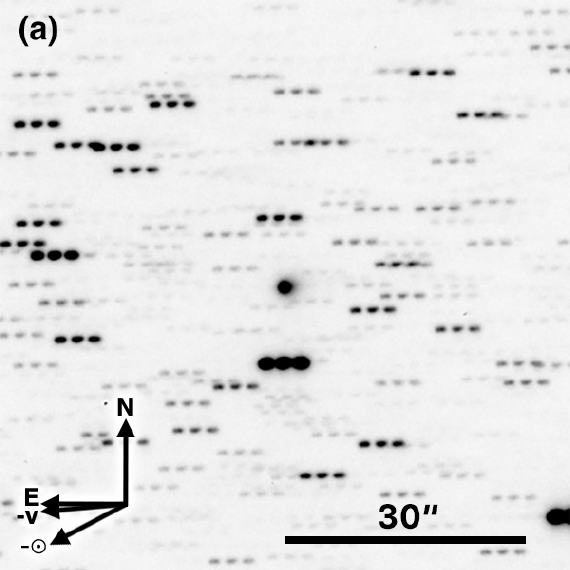}
    \includegraphics[width=1.\linewidth]{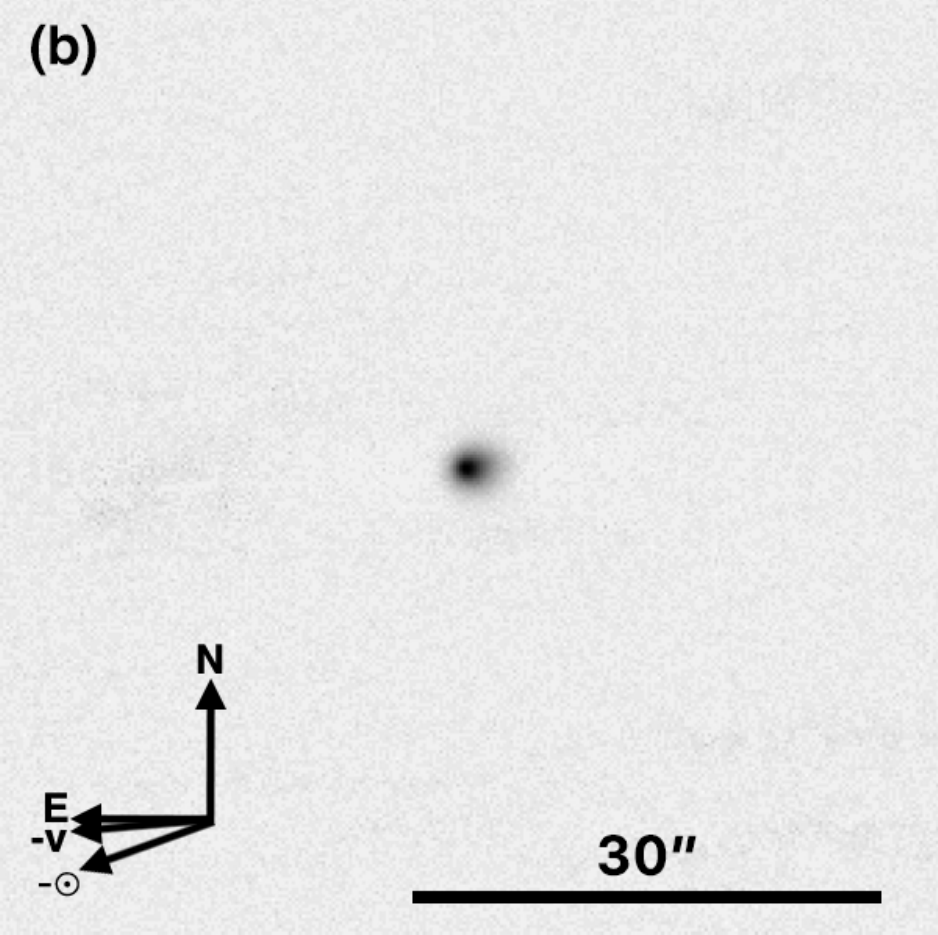}
\caption{(a) Stacked gri-band image cut-out from CFHT on 2025 July 2 and (b) R-band composite image from VLT on 2025 July 4, both showing faint activity. Background objects were removed from the VLT stack.  Arrows indicate the directions of North (N), East (E), the anti-solar vector as projected on the sky ($-\odot$), and the negative heliocentric velocity vector as projected on the sky ($-v$).}
\label{fig:cfht_vlt}
\end{figure}

Given the stark difference in visible activity levels between 1I/`Oumuamua and 2I/Borisov (Section~\ref{sec:intro}), a high priority for initial observational analysis for 3I/ATLAS was determining whether it was active or not. With this in mind, we obtained three 60~s non-sidereally guided \textit{gri}-band images with the MegaCam wide-field mosaic imager \citep{boulade2003_megacam} on the 3.6-meter Canada-France-Hawaii Telescope (CFHT, MPC code T14) on 2025 July 2 to search for faint cometary activity. The highest-quality image had typical stellar FHWMs of $0\farcs72\pm0\farcs05$ measured perpendicular to the direction of trailing, while the object had a FWHM of $1\farcs29 \pm 0\farcs02$. The magnitude in a $5\farcs0$ radius aperture was 17.2 in the \textit{Gaia} DR2 G band after three background stars were masked, although the field was crowded, so we consider this measurement to be unreliable. Figure \ref{fig:cfht_vlt}a shows the stacked composite image of these data in which faint activity is visible.

\begin{figure}
    \includegraphics[width=1.\linewidth]{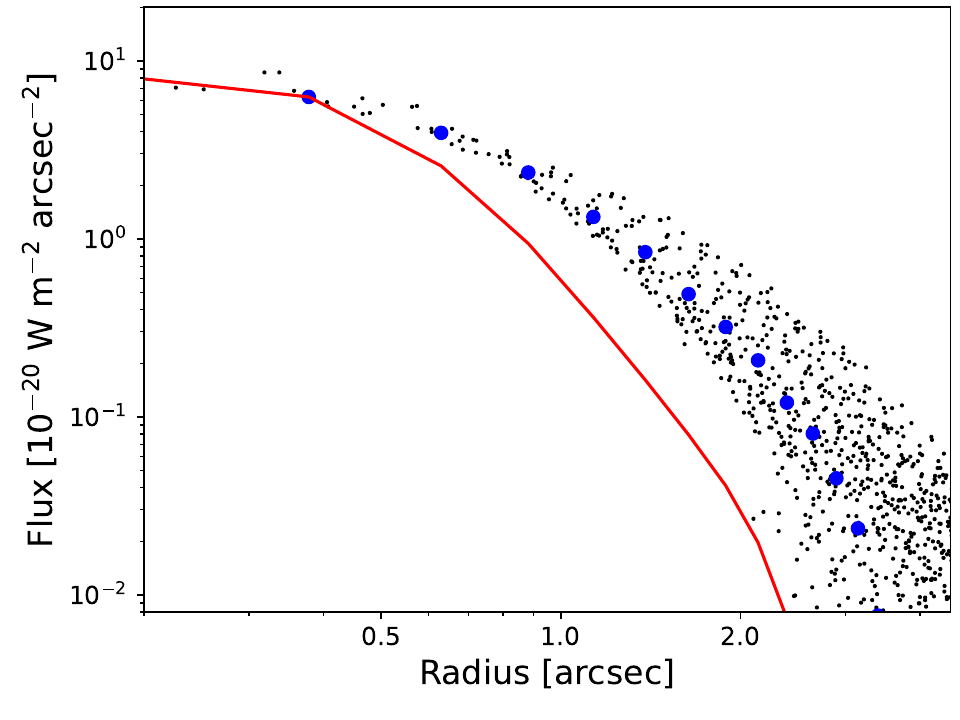}
\caption{Photometric profile of the VLT image (Figure~\ref{fig:cfht_vlt}a). The individual pixels are represented by black dots; average in annuli by blue circles. The red line is the average stellar profile measured on 10 nearby well-exposed stars.
}
\label{fig:vlt_profile}
\end{figure}

On 2025 July 4, fifteen 10~s \textit{R}-band images were acquired using the FORS2 instrument on the 8.2-meter Unit Telescope 1 of the European Southern Observatory Very Large Telescope (VLT) on Cerro Paranal, Chile. The measured stellar image quality was $0\farcs6$ FWHM. However, the object had a FWHM of $1\farcs49 \times 1\farcs29$ with the asymmetric elongation pointing to position angle $277^{\circ}$ from North to East, where the coma can be seen extending out to $3\farcs5$. A composite image of these data is shown in Figure~\ref{fig:cfht_vlt}b. The photometric profile of the object (Figure~\ref{fig:vlt_profile}) shows a brightness excess extending over $4\farcs0$ compared to that of field stars.

\begin{figure}
    \includegraphics[width=1.\linewidth]{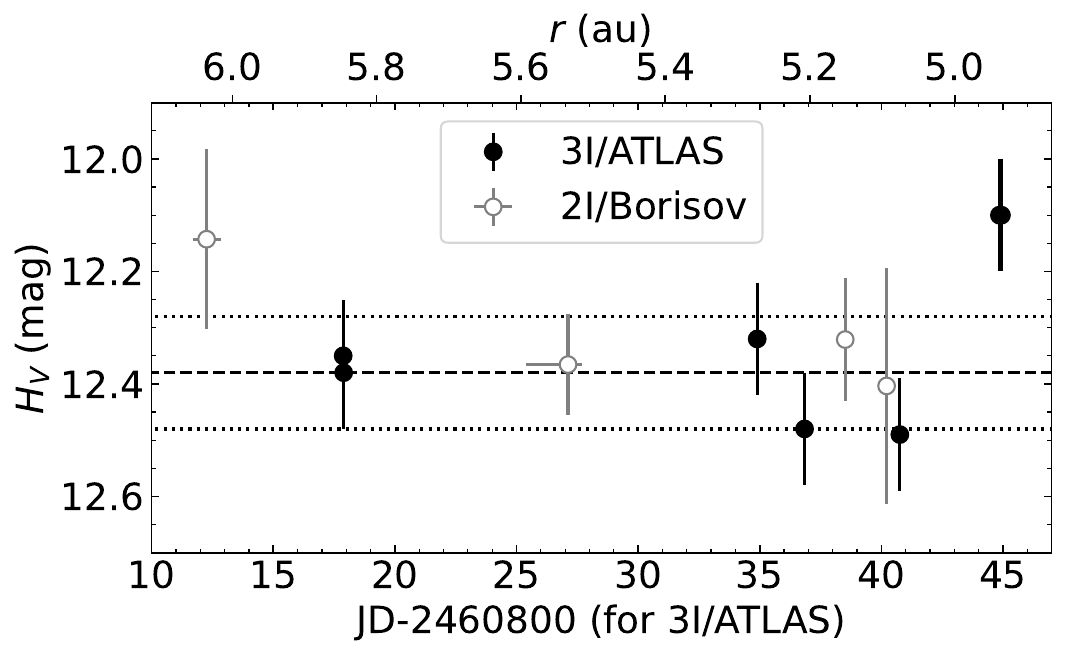}
    \caption{Absolute magnitudes computed from ZTF precovery data published in MPEC 2025-N51 from 2025 May 22 to June 18  (black filled dots).  The median $H_V$ and standard deviation of data  are shown by a dashed horizontal line and dotted horizontal lines, respectively. No uncertainties were reported for these original data, and so plotted uncertainties are estimated based on the standard deviation of photometric points measured between 2025 May 22 and June 14.   Corresponding pre-perihelion $H_V$ for 2I/Borisov derived from \citet{ye2020_borisov} are plotted on the same heliocentric distance ($r$) scale (open circles) for comparison.}
\label{fig:ztf_secular}
\end{figure}

As noted in Section~\ref{sec:orbit},  $g'$- and $r'$-band precovery data from ZTF were published in MPEC 2025-N51 following these observations. This included photometry dating back to \added{UT 2025 May 22, all of which were measured from difference images (produced by subtracting an appropriately matched template sky image from the target image) using aperture photometry utilizing apertures with radii of $5''$ (where the image pixel scale was $1''$~arcsec$^{-1}$).} These data mostly consisted of single $g'$- or $r'$-band detections on each night, but included pairs of $g'$ and $r'$-band detections on 2025 May 22 and June 18. We compute inferred $V$-band absolute magnitude from these data  assuming a standard asteroidal phase function \citep{bowell1989_astphotmodels_ast2} assuming $G=0.15$ and solar colors \citep{jordi2006_filtertransformations,holmberg2006_solarcolors}. We use this to present a secular light curve  in Figure~\ref{fig:ztf_secular}. The computed absolute magnitudes are relatively constant between 2025 May 22 and June 14 (during which the object ranged between heliocentric distances of $r=5.8$~au to $r=5.1$~au). The secular light curve has a median value of $H_V\sim12.4$ and a standard deviation of  $\sim$0.1 mag. 
\added{No uncertainties were reported for these ZTF data in MPEC 2025-N51. Therefore we  adopt this standard deviation as an approximate uncertainty level for all ZTF photometry reported here for 3I/ATLAS based on the scatter in the available photometric points between 2025 May 22 and June 14.}

\added{Figure~\ref{fig:ztf_secular} also includes the corresponding $H_V$ of 2I/Borisov's unresolved coma from pre-discovery ZTF observations of that comet \citep{ye2020_borisov} at similar $r$ for comparison. If the 3I/ATLAS photometry is also measuring the brightness of an unresolved coma, the similarity of the two interstellar comets' $H_V$ magnitudes at $r\simeq$5--6~au pre-perihelion suggests that 3I/ATLAS may have a kilometer-scale nucleus comparable in size to that of 2I/Borisov, if the two nuclei are similar in composition or otherwise eject a comparable amount of dust per unit cross section under equivalent solar heating.}

The photometry from 2025 June 18 (the last night plotted in Figure~\ref{fig:ztf_secular} when the object was at $r=4.9$~au) corresponds to an absolute magnitude outside that standard deviation at  $H_V=12.1\pm0.1$.  For comparison, $r'$-band light curve data obtained from LCO-network telescopes (discussed in more detail in Section~\ref{sec:rotation}) show an approximate average $r'$-band apparent magnitude of $m_r\sim17.9$, corresponding to $H_V=11.9$, indicating that the faint coma visible in CFHT and VLT data obtained on 2025 July 2 and 4 could be responsible for about 0.5~mag of excess brightening of the object since early June.

We note that this finding does not exclude the presence of coma prior to mid-June, but indicates that any coma that may have been present was relatively steady-state from late May to early June, and then increased in mid-June. For reference, $ H_V=12.4\pm0.1$ corresponds to a radius of $ \sim(10\pm1)$ km for  a geometric albedo of $ p_V=0.05$ typical of comets, but this is an upper limit given that we cannot rule out activity during the observations used to derive this result.

\begin{figure*}
\includegraphics[width=1.\linewidth]{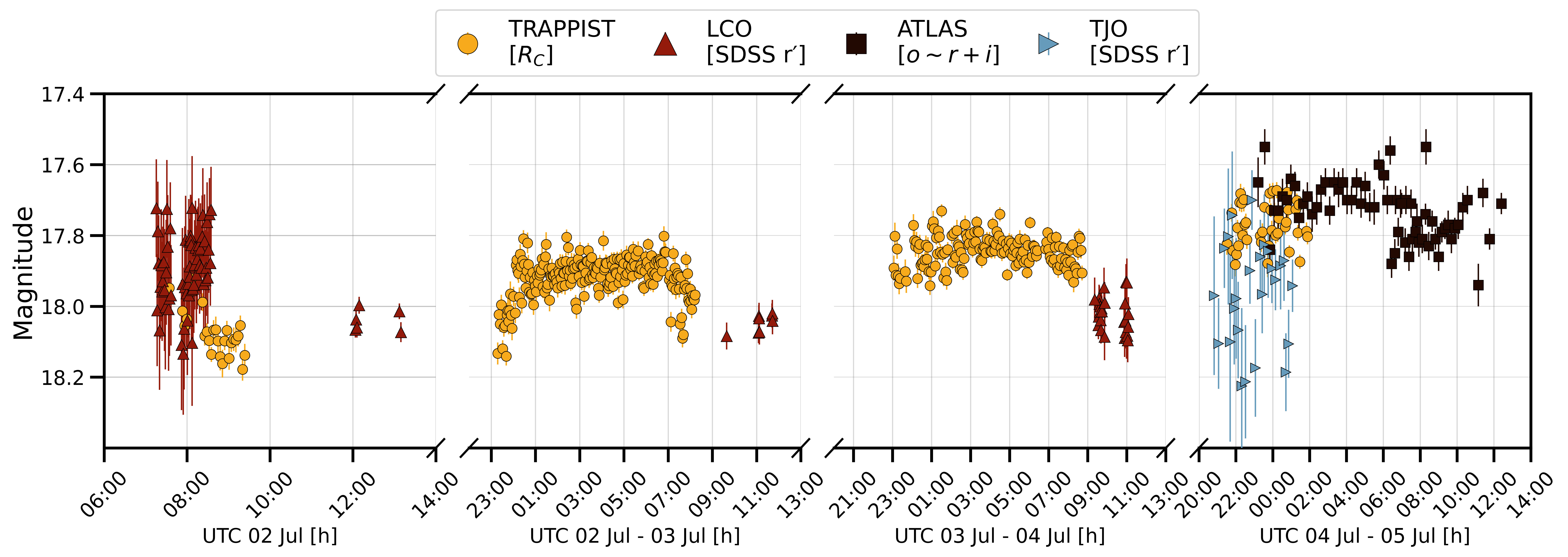}
\caption{Compiled light curve of 3I/ATLAS incorporating $r'$-band data from the LCO 0.35-meter telescopes, Faulkes Telescope North, Faulkes Telescope South, and the Telescope Joan Or\'{o}, Johnson-Cousins $Rc$-band data from the TRAPPIST-North and -South telescopes, and $o$-band data from ATLAS-HKO and ATLAS-CHL.
 A summary of parameters used in the measurement of these data is shown in Table~\ref{tab:lcfacilities}.  Tabular data for photometry plotted in this figure are available at \url{https://github.com/3I-ATLAS/discovery-paper}.
\href{https://github.com/3I-ATLAS/discovery-paper/blob/main/src/scripts/light_curve.py}{\githubicon}
}
\label{fig:lc}
\end{figure*}

\subsection{Rotation}\label{sec:rotation}

\begin{table*}
\centering
\caption{Light Curve Observations Summary}
\begin{tabular}{lcccccc}
\hline
  & MPC & Observation &  & Pixel & Photometry & Aperture \\
Facility/Instrument & Site Code & Dates (UT) & Filter(s) & Scale & Method & Radius$^a$ \\
\hline
ATLAS (Haleakala)         & W68 & 2025 July 4    & $o$ & $1\farcs86$ & PSF & n/a \\
ATLAS (Chile)             & T05 & 2025 July 4    & $o$ & $1\farcs86$ & PSF & n/a \\
Faulkes Telescope North   & F65 & 2025 July 2, 4 & $r'$  & $0\farcs28$ & aperture & $2\farcs2$ \\
Faulkes Telescope South   & E10 & 2025 July 2, 4 & $r'$  & $0\farcs28$ & aperture & $2\farcs2$ \\
LCO 0.35m (Haleakala \#1) & T03 & 2025 July 2, 4 & $r'$  & $0\farcs74$ & aperture & $3\farcs0$ \\
LCO 0.35m (Haleakala \#2) & T04 & 2025 July 2, 4 & $r'$  & $0\farcs74$ & aperture & $3\farcs0$ \\
Telescope Joan Or{\'o}    & C65 & 2025 July 4-5  & $r'$  & $0\farcs72$  & aperture & $2\farcs2$ \\
TRAPPIST-North & Z53      & 2025 July 2-4        & $R_c$ & $1\farcs20$ & aperture & $4\farcs8$ \\
TRAPPIST-South & I40      & 2025 July 2-4        & $R_c$ & $1\farcs28$ & aperture & $5\farcs1$ \\
\hline
\multicolumn{7}{l}{$^a$ Photometry aperture size if aperture photometry is used; otherwise, this field is not applicable (n/a).}
\end{tabular}
\label{tab:lcfacilities}
\end{table*}

In order to characterize 3I/ATLAS's rotational light curve, we obtained time-series observations between 2025 July 2 and July 4 using two separate 0.35-meter telescopes from the LCO Haleakal${\rm\bar{a}}$ observatory (MPC codes T03, T04), the 2.0-meter Faulkes Telescope North (FTN, MPC code F65) at Haleakal${\rm\bar{a}}$ Observatory, and the 2.0-meter Faulkes Telescope South (FTS, MPC code E10) at Siding Spring Observatory, all of which are part of the same network.  Time-series observations were also obtained on 2025 July 4--5 using the Telescope Joan Or\'{o} (TJO, MPC code C65) at Montsec Observatory.

Photometric observations \added{of 3I/ATLAS} were obtained on the nights of 2025 July 2--4 with the TRAPPIST telescopes \citep{jehin2011trappist}. TRAPPIST-North (TN, MPC code Z53) is located at the Oukaimeden observatory in Morocco, and TRAPPIST-South (TS, MPC code I40) is located at the ESO La Silla Observatory in Chile. Both telescopes are 0.6-meter Ritchey-Chretien telescopes operating at f/8. The TN camera is an Andor IKONL BEX2 DD camera providing a $20'$ field of view and pixel scale of \added{$1\farcs20$~pixel$^{-1}$ (using $2\times2$ binning)}. TS is equipped with a FLI ProLine 3041-BB CCD camera with a $22'$ field of view and a pixel scale of \added{$1\farcs28$~pixel$^{-1}$ (using $2\times2$ binning)}. Johnson-Cousin $R_c$ filters were used with individual exposure times of 180~s \added{on both telescopes}.

\added{To compensate for the dense star fields in the TRAPPIST observations, an average template image constructed from ten adjacent images is subtracted from each image in each set of data, where reference stars are used to compensate for extinction variations between science and template images. 
%and for background registration. 
This results in far less background-source-contaminated data from which photometry of 3I/ATLAS can be measured.  We then perform aperture photometry using \textsc{PhotometryPipeline} \citep{Mommert_2017} by matching $\sim$100 field stars in each image with the Pan-STARRS DR1 photometric catalog and using photometry aperture radii of 4 pixels ($4\farcs8$ on TN and $5\farcs1$ on TS).}

We also obtained time-series observations of 3I/ATLAS in $o$-band (AB; approximately equivalent to $r'+i'$) using the ATLAS network starting at $\sim$23:00 UTC on 2025 July 4, interleaving 3I/ATLAS exposures approximately every four of normal survey operations.  Using the ATLAS sites in Chile and Haleakal${\rm\bar{a}}$ (ATLAS-HKO and ATLAS-CHL; MPC codes W68 and T05), a 13 h continuous light curve was obtained from the normal differencing pipeline. \added{These photometric measurements were performed on difference images using PSF-based forced photometry (i.e., fitting a stellar PSF to the flux at the expected position of the target in each image, and measuring the magnitude of the best-fit PSF, which, it should be noted, may underreport the brightness of extended sources like 3I/ATLAS) using PSFs derived from nearby field stars \citep{Tonry2018a,smith2020_atlas}.} The zeropoint for each exposure is derived from $\sim10^5$ reference stars from Refcat2 \citep{Tonry2018b}.

\added{A summary of the facilities and photometry methods used to collect and measure these data is shown in Table~\ref{tab:lcfacilities}.} Compiled light curve results are shown in Figure~\ref{fig:lc}.  \added{In this plot, offsets between data sets may be attributed to different filters being used, differences between photometry measurement methods (i.e., PSF-based photometry for ATLAS data versus aperture photometry for all other data sets), differences in apertures used for the data sets measured using aperture photometry, and changes in observing geometry as these data have not been corrected for changes in heliocentric and geocentric distance or solar phase angle}. 

No periodicity is immediately apparent within any of the datasets, indicating little brightness variation \added{$\lesssim0.2$~mag} during the observation period\added{, where some scatter may also be due to background source contamination (or residual contamination in the cases of data for which image differencing has been applied).}
This is in notable contrast to 1I/`Oumuamua, which displayed a far more extreme light curve range of $\sim$ 3.5~magnitudes (see Section~\ref{sec:intro}). It should be noted, however, that the detection of activity and coma (Section~\ref{sec:activity}) means that any photometric variations caused by a rotating nucleus will be at least partially suppressed by the relatively more steady state coma surrounding it.  \added{A more detailed analysis of these data will be presented in a future work, though we note that \citet{Marcos2025} reported a tentative periodicity in time-series observations, finding a period of 16.79~hr, minimum brightness at MJD 60861.00, and an amplitude 0.08 magnitude, which agrees  well with our 14h ATLAS light curve.}

\subsection{Colors}\label{sec:colors}

Two multi-filter imaging sequences were obtained with the 2.0-meter Faulkes Telescope North on Haleakal${\rm\bar{a}}$ using the four-channel MuSCAT3 imager, which records the Sloan $g'$-, $r'$-, $i'$-, and $z_{\rm s}$-bands simultaneously: (i) six exposures of 30 s in each filter, and (ii) three exposures of 50~s in each filter, yielding total integrations of 180~s and 150~s per band, respectively. The data output by the LCO reduction pipelines were processed with \textsc{PhotometryPipeline} \citep{Mommert_2017}, which performs SCAMP astrometric solutions against \textit{Gaia}~DR3 \citep{gaiacollaboration2023_gaiadr3} and aperture photometry \added{ using apertures with radii of $1\farcs46$ (5.4 pixels)} and calibrated to Pan-STARRS~DR2 \citep{flewelling2020_ps1}. \added{ From these data, we find mean colors of $g'-r'=0.85\pm0.03$, $r'-i'=0.25\pm0.03$, and $i'-z'=0.20\pm0.08$.}

 \begin{figure}
\includegraphics[width=1.\linewidth]{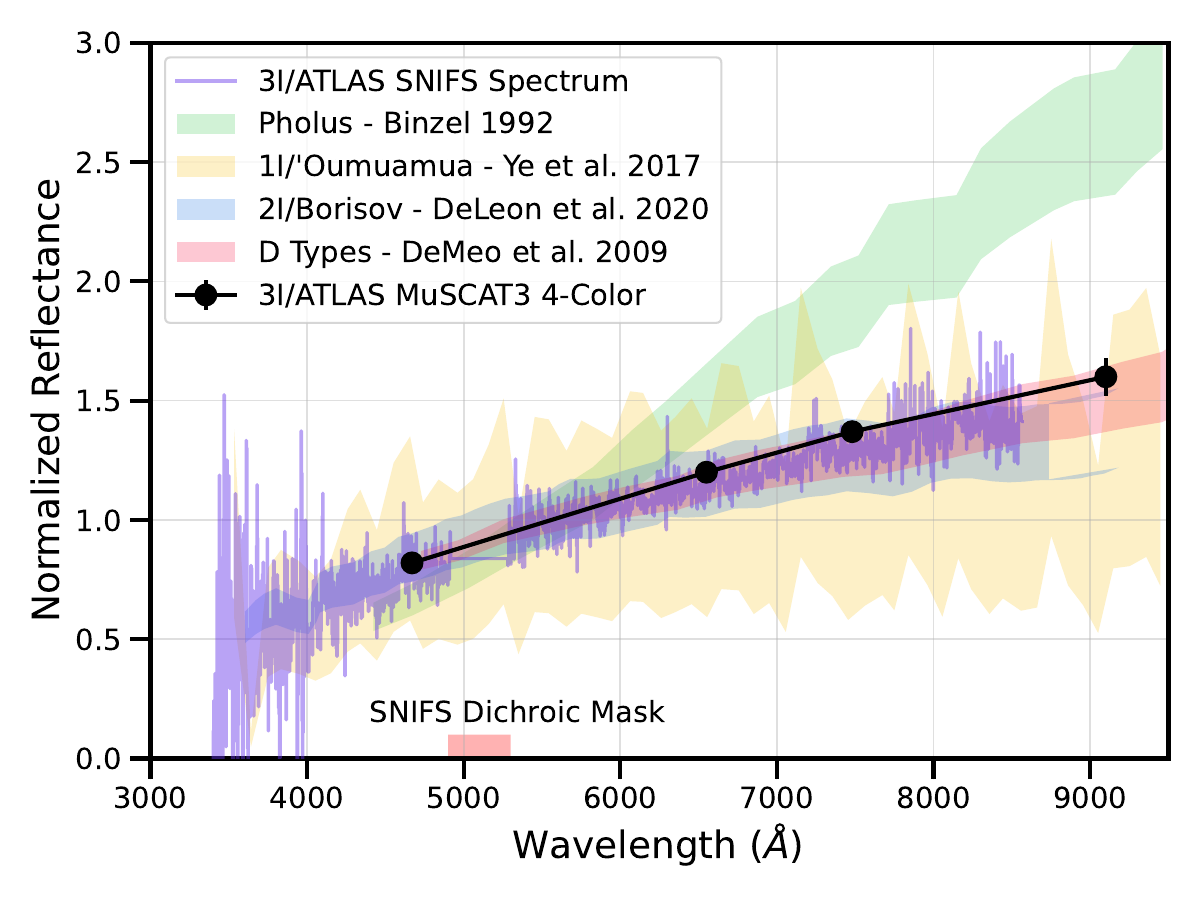}
\caption{The $g'$,$r'$,$i'$,$z'$ colors of 3I/ATLAS obtained with FTN converted to a solar reflectivity as well as the reflectance spectrum obtained with SNIFS on the UH 2.2-meter at Maunakea normalized at 5500 $\mathrm{\AA}$ are plotted in comparison to 1I/`Oumuamua \citep{Ye2017}, 2I/Borisov \citep{deleon2020}, the extremely red Centaur Pholus \citep{Binzel1992}, and the mean D-type asteroid spectrum \citep{DeMeo2009}. The errors on the FTN photometric data are approximately the size of the plot points. The region of the spectrum masked due to the SNIFS dichroic sensitivity is indicated in red. Both the observed spectrum and the four-color derived reflectance slopes are in agreement. 3I/ATLAS shows a moderately red spectral slope similar to 1I/`Oumuamua and 2I/Borisov.
\href{https://github.com/3I-ATLAS/discovery-paper/blob/main/src/scripts/3I_color_plot.py}{\githubicon}
}
\label{fig:3I_colors}
\end{figure}

\added{In Figure \ref{fig:3I_colors}, we show the resulting four-color surface reflectance spectrum of the object from the FTN.} The colors of 3I/ATLAS as seen in Figure \ref{fig:3I_colors} are relatively linear (e.g., without obvious absorption features or spectral curvature) and significantly redder than those of the Sun. This slope of this reflectance spectrum is approximately 18  \%/100 nm as derived from the full range of the available color data. It is similar, though somewhat redder, to that retrieved for 1I/`Oumuamua \citep[10$\pm$6  \%/100 nm,][]{Ye2017} and 2I/Borisov  \citep[12$\pm$1  \%/100 nm][]{deleon2020}, both of which are slightly redder than the D-type asteroids \citep{DeMeo2009}. \added{Moreover, it is within the range found for 1I/`Oumuamua of 7-23  \%/100 nm reported by \citet{Fitzsimmons2017}.} However, it is not as red as seen in some outer Solar System objects  \citep[e.g. Pholus,][]{Binzel1992}. Some of the reflected light must be from the coma because 3I/ATLAS is weakly active. This is corroborated by the fact that photometry of $g'$- and $r'$-band ZTF precovery data (Section~\ref{sec:activity}) obtained when the object appeared to be less active, indicated that the $g'-r'$ colors on those nights were close to solar  (within large uncertainties). Specifically, the measurements were $ (g'-r')=0.42\pm0.14$ on 2025 May 22 and $ (g'-r')=0.44\pm0.14$ on June 18, compared to $ (g'-r')_{\odot}=0.45\pm0.02$ \citep{holmberg2006_solarcolors}. This suggests that the much redder color measured from later data could be heavily affected by ejected dust. The spectral slope measured here is slightly higher than the average for cometary dust (see modeling and discussions of typical Solar System comets in \citealt{protopapa2018, kareta_noonan23}, as well as Figure 3 of \citealt{2024come.book..621K}), but not significantly so. 2I/Borisov's coma was similarly red (see, e.g., \citealt{deleon2020}), but the object was also significantly more active at the time of the color observations.

\subsection{Spectrum}\label{sec:spectrum}

The spectrum shown in Figure \ref{fig:3I_colors} was obtained on 2025 July 4 using the SNIFS instrument (SuperNova Integral Field Spectrograph; \citealp{Lantz2004}) on the UH 2.2-meter telescope at Maunakea by the Spectroscopic Classification of Astronomical Transients (SCAT; \citealp{Tucker2022}) team. Data reduction followed the procedures described in \citet{Tucker2022}. SNIFS contains two channels split by a dichroic mirror, with a blue channel covering 0.34--0.51~$\mu$m, and a red channel covering 0.51--1.0~$\mu$m, with spectral resolutions of 5 \r{A} and 7 \r{A}, respectively. The fast reduction pipeline used here does not include the full dichroic correction; therefore, data in this region were excluded from the final spectrum. In addition to 3I/ATLAS, the solar analog HD~165290 was observed. The final calibrated reflectance spectrum obtained by dividing the target spectrum by the normalized spectrum of the solar analog is shown in Figure~\ref{fig:3I_colors}.

The normalized reflection spectrum shows no obvious absorption features or gas emissions and yields a spectral slope of 17.1$\pm$0.2  \%/100 nm \added{measured from the spectrum window shown in Figure ~\ref{fig:3I_colors}}. This value was determined with a linear fitting routine with bootstrapping to estimate uncertainties. This spectral slope confirms the moderately red slope observed with the MuSCAT3 images, further solidifying a color similarity to both D-types and 2I/Borisov. 

\section{Discussion}\label{sec:discussion}

Given that three interstellar objects have now been discovered, it is worthwhile to use their measured physical properties and discovery circumstances to estimate the size of the total population in the Solar System.

\begin{figure}
\includegraphics[width=80mm]{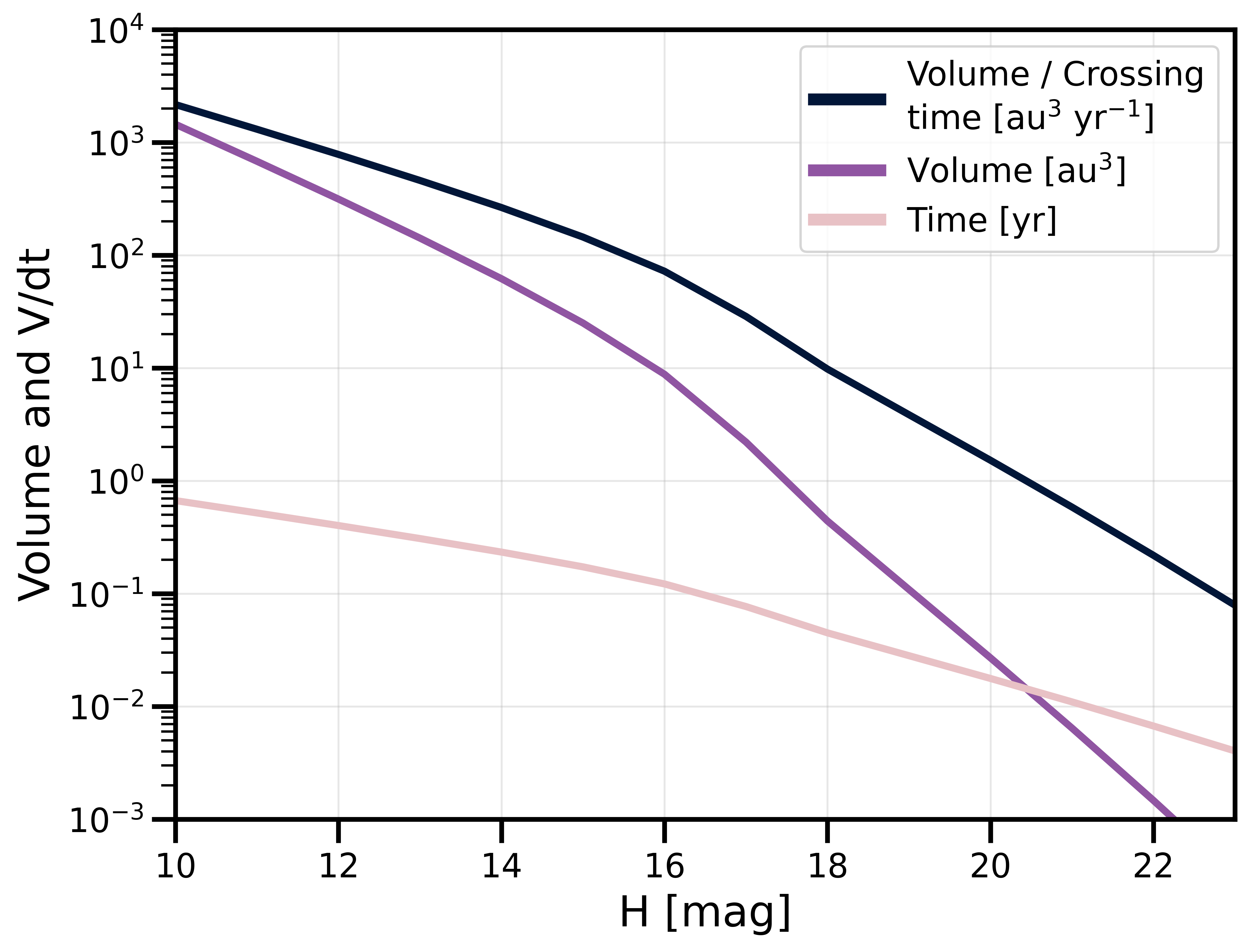}
\caption{The volume in which ATLAS can detect an object is plotted as a function of H magnitude (red).  The ratio of volume and crossing time (blue) is approximately the product of detection cross section and velocity.  When the visibility time (green) becomes less than $\sim0.1$~yr (a lunation) weather and other effects will diminish the chances of a detection below the $V/dt$ curve.
\href{https://github.com/3I-ATLAS/discovery-paper/blob/main/src/scripts/atlas_detection.py}{\githubicon}
}
\label{fig:3I_rate}
\end{figure}

The volume within which ATLAS can detect a moving object depends strongly on intrinsic brightness ($H_V$) and location in the Solar System, where sensitivity, illumination, distance, phase function, and trailing losses are critical components. Using an ATLAS detection model that assumes a $G=0.15$ phase function, a speed of 50~km~s$^{-1}$ for trailing losses, a weather and moon averaged limiting $m_{lim}\simeq19$, and 50~km~s$^{-1}$ speed for volume crossing, Figure~\ref{fig:3I_rate} shows the volume in which ATLAS can detect an object.

This curve of detectability volume divided by crossing time has a slope of $\log(V/dt)\propto -0.2H_V$ for $H_V<15$ and a slope of $\log(V/dt)\propto -0.4H_V$ for $H_V>15$.  However, the crossing time for $H>16$ is less than 0.1~yr which means that the actual detectability is certainly less than $V/dt$ from Figure~\ref{fig:3I_rate} because of weather and other survey cadence effects. The ratio of detectability volume and crossing time is approximately the product of detection cross section and velocity, hence the ratio of detection rate and local density.  Evaluating Figure~\ref{fig:3I_rate} at $H_V\sim12.5$ for 3I/ATLAS gives $600$~au$^3$~yr$^{-1}$, and a local density of $\sim3\times10^{-4}$~au$^{-3}$ for a detection rate of $\sim$ 0.2~yr$^{-1}$. \added{While this calculation assumes a slightly different magnitude than that measured for 3I/ATLAS, it is worth noting that $H_V$ will likely change for 3I/ATLAS depending on whether the activity level changes. In this case since $H_V$ of 3I/ATLAS was measured when active this limiting search volume corresponds to the absolute brightness of the inert nucleus.}

The $G=0.15$ phase function used for Figure~\ref{fig:3I_rate} may be inappropriate for an object such as 3I/ATLAS with an unknown amount of coma, and unknown and potentially unfamiliar surface scattering properties, which could cause this density estimate to be significantly incorrect.  Obtaining stronger constraints on 3I/ATLAS's nucleus size, such as with high-resolution imaging by space telescopes or adaptive optics-equipped ground-based telescopes, or by searching for occultation observation opportunities, should therefore be considered a high priority.

\cite{Do2018} estimated an interstellar density of objects like 1I/`Oumuamua at 0.2~au$^{-3}$ ($H_V\sim23$) and noted that the size distribution must be steeper than $n(d>D)\propto D^{-3}$ to avoid a logarithmic divergence in total mass, i.e. differential $\log n(d)$ with slope greater than $0.8H_V$ \citep[see Fig. 3 of ][]{Do2018}.  This criterion applied to the ATLAS detectability curve would cause a strong peak around $H\sim16$, not two detections at $H\sim13$, however.  As such, the detections of 2I/Borisov and 3I/ATLAS, both of which appear, or appeared, relatively large, are somewhat unexpected given that more numerous smaller interstellar objects have not also been discovered over the same time period.  Clearly what is needed is more interstellar object discoveries to improve population statistics, which is hopefully exactly what the Rubin Observatory Legacy Survey of Space and Time \citep[LSST;][]{jones2009_lsst} will deliver in the coming years.

We are certain to learn more about 3I/ATLAS in the coming weeks and months as it experiences continued and increasing heating for perhaps the first time during its passage past our Sun. Comprehensive and collaborative investigations of 3I/ATLAS based off the lessons learned from the 1I/`Oumuamua and 2I/Borisov campaigns are poised to significantly expand our knowledge of the interstellar object population, and as such, additional observations of 3I/ATLAS are highly encouraged.  Further photometry, spectroscopy, or polarimetry could constrain the rotation, activity, dust size-frequency distribution, coma composition, and nongravitational acceleration of the object. The perihelion of 3I/ATLAS will not be easily observable from Earth-based observatories, as the object will be on the opposite side of the Sun and at a low solar elongation angle. However, the object will approach within 0.19 au of Mars, and we encourage nearby spacecraft (e.g., Mars Reconnaissance Orbiter, Trace Gas Orbiter, MAVEN, Tianwen-1, Hope, Jupiter Icy Moons Explorer) equipped with visible, UV, and IR spectrographs and cameras to attempt to capture data on this object's closest approach.  

\section{Data and Software Availability\label{section:software}}

Data behind many of the figures in this work are available at \url{https://github.com/3I-ATLAS/discovery-paper}.

\section{acknowledgments}

\added{We thank the anonymous reviewer for insightful comments and constructive suggestions that strengthened the scientific content of this manuscript. We thank the scientific editor Dr. Edgard G. Rivera-Valent\'in for obtaining reports in a timely manner. }

D.Z.S.\ is supported by an NSF Astronomy and Astrophysics Postdoctoral Fellowship under award AST-2303553. This research award is partially funded by a generous gift of Charles Simonyi to the NSF Division of Astronomical Sciences. The award is made in recognition of significant contributions to Rubin Observatory's Legacy Survey of Space and Time. \added{A.Y. and T.F. also acknowledge support from NSF grant number AST-2303553.}
D.F.\ conducted this research at the Jet Propulsion Laboratory, California Institute of Technology, under a contract with the National Aeronautics and Space Administration (80NM0018D0004).
T.S-R.\ acknowledges funding from Ministerio de Ciencia e Innovaci{\'o}n (Spanish Government), PGC2021, PID2021-125883NB-C21. This work was (partially) supported by the Spanish MICIN/AEI/10.13039/501100011033 and by ``ERDF A way of making Europe" by the “European Union” through grant PID2021-122842OB-C21, and the Institute of Cosmos Sciences University of Barcelona (ICCUB, Unidad de Excelencia `Mar\'ia de Maeztu’) through grant CEX2019-000918-M. A.D.F. acknowledges funding from NASA through the NASA Hubble Fellowship grant HST-HF2-51530.001-A awarded by STScI. K.J.M., J.W., and A.H.\ acknowledge support from the Simons Foundation through SFI-PD-Pivot Mentor-00009672. A.G.T.\ acknowledges support from the Fannie and John Hertz Foundation and the University of Michigan's Rackham Merit Fellowship Program. D.M.\ acknowledges support by the Science Fund of the Republic of Serbia, GRANT No.\ 7453, Demystifying enigmatic visitors of the near-Earth region (ENIGMA). E.P.-A.\ acknowledges support by the Italian Space Agency within the LUMIO project (ASI-PoliMi agreement n.\ 2024-6-HH.0).  B.J.S, K.H., and W.B.H. acknowledges support from NSF (grants AST-2407205) and NASA (grants HST-GO-17087, 80NSSC24K0521, 80NSSC24K0490, 80NSSC23K1431).  W.B.H acknowledges support from the National Science Foundation Graduate Research Fellowship Program under Grant Nos. 1842402 and 2236415. \added{Q.Y. is supported by NASA grant 80NSSC21K0659.}

This material is based upon work supported by the National Science Foundation Graduate Research Fellowship Program under Grant Nos.\ 1842402 and 2236415. Any opinions, findings, conclusions, or recommendations expressed in this material are those of the author(s) and do not necessarily reflect the views of the National Science Foundation. K.E.M. acknowledges support from the NASA ROSES DDAP program funded through NASA Goddard Space Flight Center.

This work has made use of data from the Asteroid Terrestrial-impact Last Alert System (ATLAS) project. ATLAS is primarily funded to search for near-Earth asteroids through NASA grants NN12AR55G, 80NSSC18K0284, and 80NSSC18K1575; byproducts of the NEO search include images and catalogs from the survey area.  The ATLAS science products have been  made possible through the contributions of the University of Hawaii Institute for Astronomy, the Queen's University Belfast, the Space Telescope Science Institute, the South African Astronomical Observatory (SAAO), and the Millennium Institute of Astrophysics (MAS), Chile.

This work also includes observations obtained with MegaPrime/MegaCam, a joint project of CFHT and CEA/DAPNIA, at the Canada-France-Hawaii Telescope (CFHT) which is operated by the National Research Council (NRC) of Canada, the Institut National des Science de l'Univers of the Centre National de la Recherche Scientifique (CNRS) of France, and the University of Hawaii. The observations at the Canada-France-Hawaii Telescope were performed with care and respect from the summit of Maunakea which is a significant cultural and historic site.

This paper is based on observations made with the MuSCAT3 instrument, developed by Astrobiology Center and under financial supports by JSPS KAKENHI (JP18H05439) and JST PRESTO (JPMJPR1775), at Faulkes Telescope North (FTN) on Maui, Hawaii, operated by the Las Cumbres Observatory.  Coordination of some observations conducted using LCO network telescopes was carried out as part of the LCO Outbursting Objects Key project \citep[LOOK;][]{lister2022_look}.

The Joan Or\'{o} Telescope (TJO) at the Montsec Observatory (OdM) is owned by the Catalan Government and operated by the Institute of Space Studies of Catalonia (IEEC).

This research has made use of data and/or services provided by the International Astronomical Union's Minor Planet Center.

\bibliography{sample7}{}
\bibliographystyle{aasjournal}

\end{CJK*}
\end{document}